# Internet of Things (IoT) based ECG System for Rural Health Care

Md. Obaidur Rahman[1]
Department of CSE, DUET, EUB, Gabtoli, Dhaka, Bangladesh

Mohammod Abul Kashem[2]
Department of CSE, DUET
Gazipur, Dhaka, Bangladesh

Al-Akhir Nayan[3]
Department of CSE, EUB
Gabtoli, Dhaka
Bangladesh

Most. Fahmida Akter[4]
Department of CSE, BUP
Mirpur, Dhaka, Bangladesh

Fazly Rabbi[5]
Department of Statistics
JNU, Savar, Dhaka, Bangladesh

Marzia Ahmed[6]
Department of Software Engineering
Daffodil International University
Dhaka, Bangladesh

Mohammad Asaduzzaman[7]
Department of ME, DUET
Gazipur, Dhaka, Bangladesh

*Abstract*—Nearly 30% of the people in the rural areas of Bangladesh are below the poverty level. Moreover, due to the unavailability of modernized healthcare-related technology, nursing and diagnosis facilities are limited for rural people. Therefore, rural people are deprived of proper healthcare. In this perspective, modern technology can be facilitated to mitigate their health problems. ECG sensing tools are interfaced with the human chest, and requisite cardiovascular data is collected through an IoT device. These data are stored in the cloud incorporates with the MQTT and HTTP servers. An innovative IoT-based method for ECG monitoring systems on cardiovascular or heart patients has been suggested in this study. The ECG signal parameters P, Q, R, S, T are collected, pre-processed, and predicted to monitor the cardiovascular conditions for further health management. The machine learning algorithm is used to determine the significance of ECG signal parameters and error rate. The logistic regression model fitted the better agreements between the train and test data. The prediction has been performed to determine the variation of PQRST quality and its suitability in the ECG Monitoring System. Considering the values of quality parameters, satisfactory results are obtained. The proposed IoT-based ECG system reduces the health care cost and complexity of cardiovascular diseases in the future.

## I. INTRODUCTION

IoT concept can be utilized in versatile areas such as intelligent health care system, intelligent agriculture, environmental impact predictions, automation industries, etc. [1, 2, 3, 4]. World Health Organization (WHO) [5] mentions that the uncertainty of health conditions is a widespread problem for aged people. Aged people need to check their health conditions very frequently, especially for senior cardiovascular patients. Existing cardiovascular diagnosis systems need to be improved, including modern technology to detect the heart condition in a low cost, accurate and timely manner [6, 7]. Considering the heart-related issue, electrocardiogram (ECG) monitoring used extensively in rural hospitals and health research centers [8].

Rather than IoT, Cyber-Physical systems (CPS) can be considered as data-centric technology. CPS integrates innovative functionality processes that facilitate communication, computation, and control through IoT [9]. Moreover, it contributes to an advanced intelligence system that significantly affects social life [10]. CPS concept can be activated through Micro Electromechanical Systems for networking in monitoring, computing, and controlling the physical world.

The work concentrates on developing a portable heart monitoring system in the corporation of Electrocardiogram (ECG) technology [11]. Three heart rate detection sensors are utilized to make the device that generates analog data from ECG signals. Moreover, the analog data are converted into CSV format by Arduino microcontroller. The collected data is transmitted to the cloud through a local server, processed into P, Q, R, S, T ECG parameters, and analyzed by the machine learning algorithm. The ECG monitoring process detects ECG signals incorporating non-intrusive sensors, and the signal obtained from sensors transmits through the smartphone by wireless transmission methods, such as Zigbee or Bluetooth [12, 13]. Steady heart rate detection and an immediate heartbeat monitoring system are viable parameters for heart disease. Experimentation indicated that cardiovascular disease could be treated, controlled, and prevented in a steady data monitoring process using ECG signals [14, 15].

A wearable ECG monitoring gadget-based system has been designed to associate with IoT and cloud service architecture for monitoring heart disease in this study. The sensors placed in the human chest records different ECG data through Arduino, and these data are transmitted to the IoT cloud without any delays. The HTTP and MQTT servers are put in the IoT cloud to provide users with quick and timely access to ECG data. The acquired data is stored in a non-relational database that can continuously improve data storage velocity and flexibility. In addition, a graphical user interface, accessible via the internet, is developed for the availability of cardiovascular-based data for medical experts to analyze the patient's heart conditions. The proposed IoT cloud-related systems in this work ensure the effectiveness, reliability, and accuracy of the data collected.

In addition to that, the measured heartbeat and ECG report of the patient obtained from this intelligent information system can be sent quickly through the text message, web server, and mobile application. The LIVE monitoring option of the webserver can be used by nurses and patient relatives for emergency cases. This process can be used for rural people at an affordable cost.

The features of the system are below:

- It will be a handy tool because it displays all the data and information collected solely from the internet. As a result, it minimizes the stress and pressure of the patient's relatives who work outside the home.

- Doctors can increase diagnostic accuracy by being connected to the health care system via IoT since they have all the relevant patient data at their fingertips. In a nutshell, it allows for continuous and remote patient monitoring.

- Doctors and family members may conduct their jobs without fear because they can track the patient's health status from anywhere. It also gives alerts anytime when a specific health parameter exceeds the optimal limit. Additionally, by getting a warning, doctors and family members can take the appropriate action and saves lives in the emergency.

- Rather than visiting or spending time at hospitals, most older people choose to stay at home with their loved ones. But, on the other hand, people suffer from a variety of diseases because of their stressful lifestyles and become very weak at old age. Furthermore, this project will benefit ICU patients.

## II. Materials and Methods

### A. Proposed System

The following components are used to run the project:

1) Arduino MEGA ATmega2560
2) Sensors
    *a)* ECG AD8232
    *b)* Heartbeat sensor MAX30100
3) Wi-Fi module ESP8266 A1 Cloud Inside
4) Jumper wires
5) Breadboard
6) Laptop/ computer.

In the middle of the devices, an ECG monitoring system is used. First, the patient will press their finger against the heartbeat sensor, and the IR sensor's ray will count the beats from the blood flow. Then, the H-Beat button will be pushed and waited for 20 seconds after counting the beats from the blood flow. Finally, the outcome will be transmitted, and the heartbeat rate will be displayed on a laptop screen or a mobile screen utilizing a web page or a mobile app. This complete process is depicted in Fig. 1. Next, the ECG sensor will be affixed to the patient's chest, and the patient will press the 'ECG' button to activate the function. Therefore, the ECG graph, signal, or digits will be generated.

### B. Electrical Components Control Unit

An Arduino Mega 2560 microcontroller board [16, 17] is utilized to complete the project. A 16 MHz crystal oscillator, 54 digital input/output pins, 4 UARTs, 16 analog inputs, a USB connection, an ICSP header, a power jack, and a reset button are included on the board. The ATmega2560 board is powered through a USB or an external power supply. The source of energy is automatically selected. Two power adapters are used. One is a 9v battery, and the other one is a 5v laptop USB. Heartbeats, ECG, and Wi-Fi sensors are getting 3.3v from those power sources. If more power is driven in the sensors, they may get damaged. Registers are used to minimize the source voltage and avoid damages.

### C. Hardware and Software Implementation

Arduino is the system's central control unit. A heartbeat sensor, an ECG sensor, and various manual buttons are available on the input side. The Arduino com port displays the output. The Wi-Fi Module allows to transfer of data to the cloud, and once the data is uploaded, the results can be reviewed by signing in to the server using a computer or smartphone. The block diagram (Fig. 2) describes the complete hardware process.

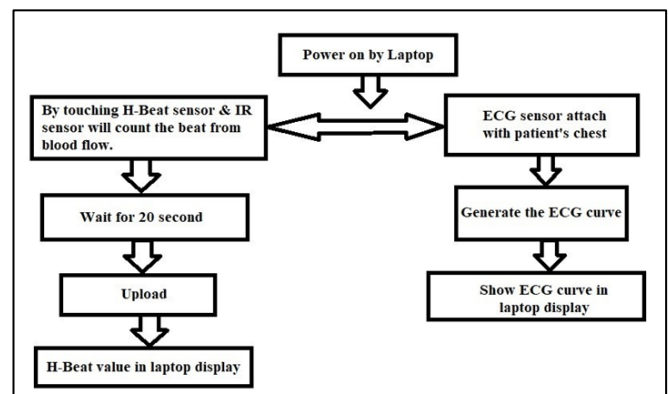

Fig. 1. Working Flow Diagram.

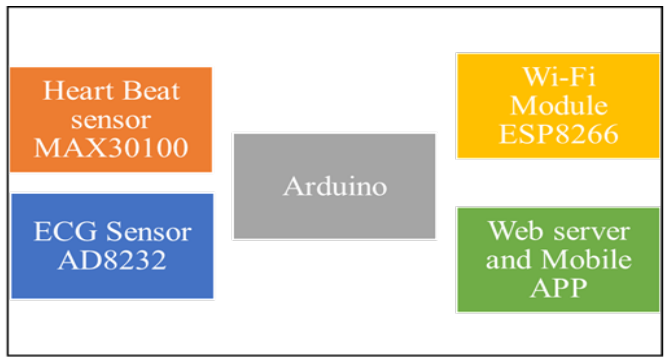

Fig. 2. Design of the IoT Device-based ECG System.

The software and hardware parts are merged in software implementation. Necessary code was written to command the hardware. In the coding section, the Arduino pin 2 (SDA and SDL) is used to initialize the cardiac bit sensor pin. Two other buttons are initialized in pin GND and 3.3v. ECG sensors LO+ and LO- are initialized in Arduino pin 10, 11. Additional controls are initialized in pin OUTPUT to A0, GND to GND, and 3.3v to 3.3v. The ECG and Heartbeat values are both zero at the beginning. The Wi-Fi cloud module contains an auto-configuration method. Wi-Fi modules RX and TX are initialized in Arduino pin 9 and 8, CH_EN and VCC are connected in Arduino pin 3.3v. Another pin named GPIO_0 is used for uploading the program to the Wi-Fi module.

In the beginning, combining two different sensors, programs are launched simultaneously and get the output result in the COM3 port of the Arduino. In the heartbeat approach, the hardware is employed to read the heartbeat. First, the heartbeat is counted for 20 seconds. At which point the heart bit pin raises high, the value is calculated. While taking data, the heartbeat is estimated for 60 seconds or 1 minute. Lastly, the heartbeat data is uploaded to the cloud. The command line for the ECG method is the same as the heartbeat method. If the ECG count is less than 50, then count the ECGs and send analog or digital data to the cloud, which is subsequently sent to the MQTT web server and displays the ECG result. The analog ECG data is uploaded and converted to digital data using the loop approach. ECG data is uploaded when it scores greater than 80, and for a lower score, an ERROR message is shown.

Wi-Fi module ESP8266 is connected to Arduino for uploading the MQTT program [18]. ESP8266 GPI 03 pin is connected to GND. After uploading the program, the RX, TX, and GPIO_0, the pin is unplugged. The RX and Tx pin of ESP8266 is connected to TX pin 8 and RX pin 9 of Arduino. After doing all these procedures, the project runs, and it transmits the sensors data to the cloud. The cloud visualizes the result on the MQTT box.

*D. MQTT Box Site Implementation*

REST has proven to be one of the most important advancements for Web applications [19]. REST stands for Representational State Transfer, and it is a technique of composing hypermedia applications. Structures for constructing RESTful Web administrations are currently included in every significant development dialect. The hardware of the project is related to a program to implement and visualize the website. On the webpage, it shows the current value. A web front interface was used in the MQTT box. For this purpose, we downloaded the MQTT box and installed the software. When the heartbeat is measured, data is uploaded into the cloud server, and then it transmits the result to the MQTT box to show real value on the website according to date and time. The ECG result is shown using a web service that calls data from the server and generates it using Arduino code.

To design and test the MQTT communication protocol, a developer's helper application was used. MQTT Box for ECG Monitoring System boosts the MQTT process. Apps for Mac, Linux, and Windows are also available for MQTT Box. In the ISM database, the cyber-level serves as the significant information hub [20]. To deploy cyberspace, information is extracted from each source and compiled. With massive data, a specific analyzer is employed to extract data that gives a better sense of each patient's ECG data monitoring status.

III. EXPERIMENTAL ANALYSIS AND RESULTS

*A. Dataset Configuration and Analysis*

We created a dataset to train the Gradient Boosting Model (GBM) [21, 22, 23]. Approximately 2000 samples were collected from volunteers. In addition, related information like age, 'P,' 'Q,' 'R,' 'S' and 'T' were also collected. Table 1 shows a tiny part of our dataset.

Exploring the relationship among different variables, some concise analyses were done to measure the mean, median, count, minimum, maximum, standard deviation, and quartile value of the dataset, which have been described in Table 2.

TABLE I. DATASET CONFIGURATION

| Record No | Age | P | Q | R | S | T |
|---|---|---|---|---|---|---|
| 1 | 21 | 91.6 | 100.0 | 100.0 | 100.0 | 90.0 |
| 2 | 23 | 100.0 | 100.0 | 100.0 | 100.0 | 100.0 |
| 3 | 30 | 100.0 | 100.0 | 100.0 | 100.0 | 100.0 |
| 4 | 25 | 100.0 | 100.0 | 100.0 | 100.0 | 100.0 |
| 5 | 20 | 78.5 | 100.0 | 80.0 | 100.0 | 100.0 |

TABLE II. DATASET MEASUREMENTS

| | R. N | Age | P | Q | R | S | T |
|---|---|---|---|---|---|---|---|
| **Count** | 20.00 | 20.00 | 20.00 | 20.00 | 20.00 | 20.00 | 20.00 |
| **Mean** | 10.50 | 29.85 | 97.06 | 96.26 | 95.26 | 96.51 | 97.66 |
| **Std** | 5.92 | 8.86 | 5.88 | 7.57 | 8.33 | 7.15 | 4.50 |
| **Min** | 1.00 | 18.00 | 78.50 | 76.19 | 76.19 | 76.19 | 85.00 |
| **25%** | 5.75 | 22.75 | 98.43 | 98.43 | 93.53 | 98.43 | 98.63 |
| **50%** | 10.50 | 28.50 | 100.00 | 100.00 | 100.00 | 100.00 | 100.00 |
| **75%** | 15.25 | 36.25 | 100.00 | 100.00 | 100.00 | 100.00 | 100.00 |
| **Max** | 20.00 | 45.00 | 100.00 | 100.00 | 100.00 | 100.00 | 100.00 |

## B. Exploring Attributes type Cleaning the Dataset

The model used the parameters described in Table 3.

Drawing Box Plot, many attributes have been found containing outliers or extreme values (Fig. 3). All the observations along with outliers were deleted. After that, some robust statistics were utilized to calculate the different features of this dataset. However, the entire dataset was considered to estimate various measures. The dataset contained no null values and needed not to treat with any null values.

## C. Dependent and Independent Variables

Quality is dependent on the target variable, and all other variables are independent variables. From these target values, a heart condition was measured (Table 4). There are six distinct quality values through analysis, and these are 14, 1, 1, 1, 2, 1. Grouping the quality, we found that most of the heart is in 14, 2 & 1, indicating the condition as good. We found two lousy quality hearts in our dataset, and three people were close to a tragic situation.

TABLE III. DATASET PARAMETERS

| S.N. | Column | Non-Null Count | Data Type |
|---|---|---|---|
| 1. | Record No | 20 not_null | int64 |
| 2. | Age | 20 not_null | int64 |
| 3. | P | 20 not_null | float64 |
| 4. | Q | 20 not_null | float64 |
| 5. | R | 20 not_null | float64 |
| 6. | S | 20 not_null | float64 |
| 7. | T | 20 not_null | float64 |

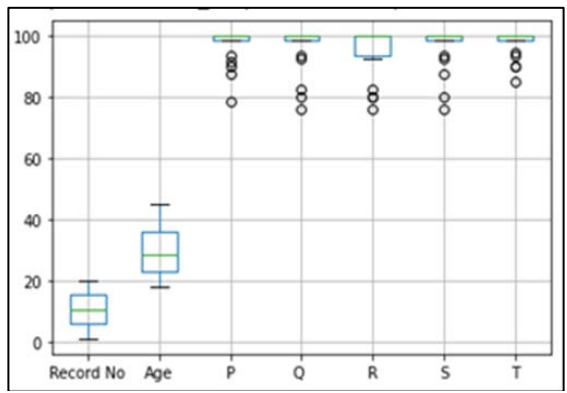

Fig. 3. BOX Plot of Dataset.

TABLE IV. QUALITY VALUES FOR GOOD OR BAD HEART CONDITION

| Mean value | Quality value |
|---|---|
| 76.13 | 1 |
| 80.00 | 2 |
| 82.35 | 1 |
| 92.85 | 1 |
| 93.75 | 1 |
| 100.00 | 14 |

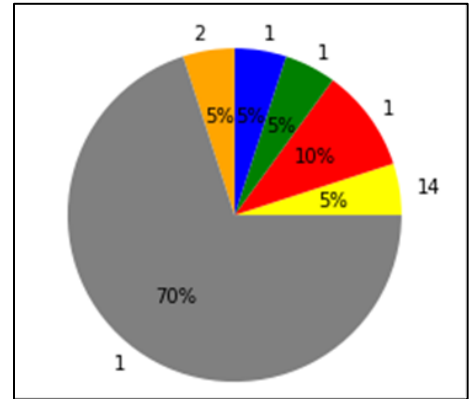

Fig. 4. Quality Wise Pie Chart.

Fig. 5. Investigation of Variables Correlation and Covariance

## D. Dataset Visualization

A quality-wise Pie-Chart was drawn that describes the percentage ratio of the six qualities. In the chart, yellow indicates quality 14, red, green, blue, and orange indicate quality 1 and 2. According to the values mentioned in Fig. 4, it can be concluded that 70.00% sample's heart condition was excellent when collecting data.

Another measurement was run to investigate the correlation and covariance of different variables with the target variables. The outcomes identify variables responsible for various heart conditions. The correlation result is mentioned in Table 5 and covariance in Table 6.

After sorting the significant correlations, we found that 'Q' (ECG, Q - parameter), 'S' (S - parameter), 'T' (T - parameter) & 'P' (P - parameter) have some moderate correlation, comparing with other attributes. On the other hand, 'R' (R – parameter included angina) has a significant positive correlation with Q (Q – parameter included angina).

TABLE V. CORRELATION VARIABLES

|  | R. N | Age | P | Q | R | S | T |
|---|---|---|---|---|---|---|---|
| R. N. | 1.0000 | 0.4431 | 0.1777 | 0.0657 | 0.1772 | 0.0408 | 0.2775 |
| Age | 0.4431 | 1.0000 | 0.4338 | 0.1623 | 0.2878 | 0.1383 | 0.1095 |
| P | 0.1777 | 0.4338 | 1.0000 | -0.2135 | 0.2052 | -0.2072 | 0.1713 |
| Q | 0.0657 | 0.1623 | -0.2135 | 1.0000 | 0.8460 | 0.9893 | 0.4546 |
| R | 0.1772 | 0.2878 | 0.2052 | 0.8460 | 1.0000 | 0.8372 | 0.3474 |
| S | 0.0408 | 0.1383 | -0.2072 | 0.9893 | 0.8372 | 1.0000 | 0.5011 |
| T | 0.2775 | 0.1095 | 0.1713 | 0.4546 | 0.3474 | 0.5011 | 1.0000 |

TABLE VI. COVARIANCE VARIABLE

|  | R. N | Age | P | Q | R | S | T |
|---|---|---|---|---|---|---|---|
| **R. N.** | 35.000 | 23.236 | 6.1802 | 2.945 | 8.735 | 1.726 | 7.389 |
| **Age** | 3.236 | 78.555 | 22.594 | 10.888 | 21.256 | 8.760 | 4.371 |
| **P** | 6.180 | 22.594 | 34.533 | -9.498 | 10.046 | -8.703 | 4.530 |
| **Q** | 2.945 | 10.888 | -9.498 | 57.285 | 53.345 | 53.515 | 15.485 |
| **R** | 8.735 | 21.256 | 10.046 | 53.345 | 69.405 | 49.846 | 13.027 |
| **S** | 1.726 | 8.7607 | -8.703 | 53.515 | 49.846 | 51.072 | 16.118 |
| **T** | 7.389 | 4.3712 | 4.530 | 15.485 | 13.027 | 16.118 | 20.251 |

### E. Comparison among Significant Variables and Target Variables

Comparing among significant variables and 'R' variables, the relationship between the variables can be virtualized. The correlation of the 'R' variable with other variables is shown in Table 7. 'Q' type makes an essential factor for having heart disease because most cases with 'R' 1 have chest pain. Two 'R' variables raise to level 120 – 150 among the age range 40 – 60. Fig. 5 represents the change of various parameters. From the findings, it can be concluded that people who have heart disease have maximum high blood pressure, and cholesterol is very high. Par's plot shows the distribution of single variables depicted in Fig. 6.

TABLE VII. CORRELATION WITH OTHER VARIABLES

| R | 1.000000 |
|---|---|
| Q | 0.846016 |
| S | 0.837237 |
| T | 0.347479 |
| Age | 0.287883 |
| P | 0.205213 |

### F. Heartbeat Result Analysis

The heartbeat result was analyzed by an automatic blood pressure system to see if the heartbeat sensor is functional or not (Fig. 7). For further processing, data were collected from five different persons of a specified age range. The information was listed along with a particular day and time.

### G. ECG Report Analysis

Three electrodes are inserted on the patient's chest at first. The red electrode is implanted on the right side of the chest. The green electrode is situated on the left side of the chest, whereas the yellow electrode is located under the green electrode. Then the ECG push button is pressed. The value is converted into a curve and uploaded to the webserver and virtualized through mobile app and website. Arduino com port result is shown in Fig. 8. The key feature of the measured ECG is depicted in Fig. 9. ECG_PQRST data from the device is mentioned in Table 8.

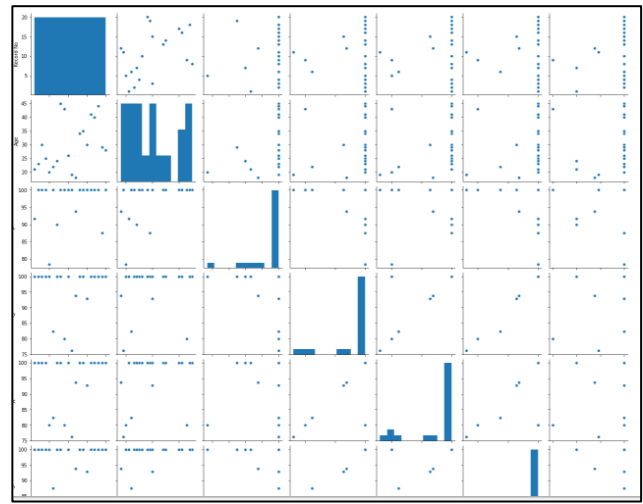

Fig. 7. Pair Plots among the Variables.

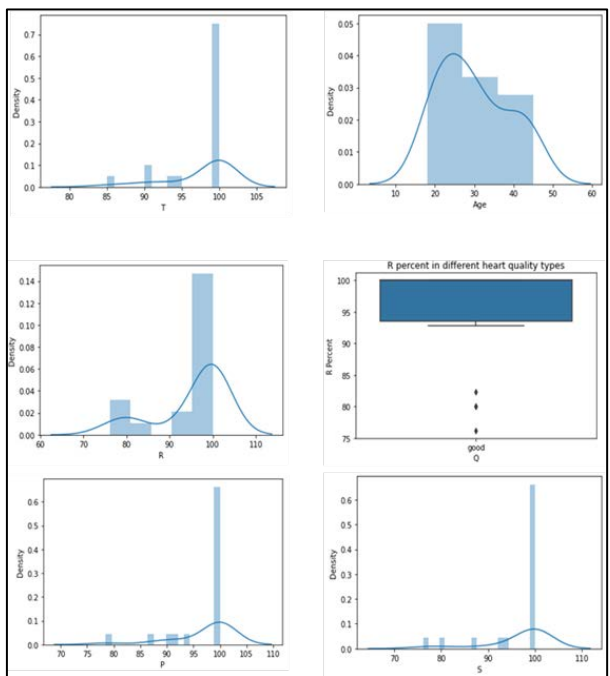

Fig. 6. Change of Various Parameters.

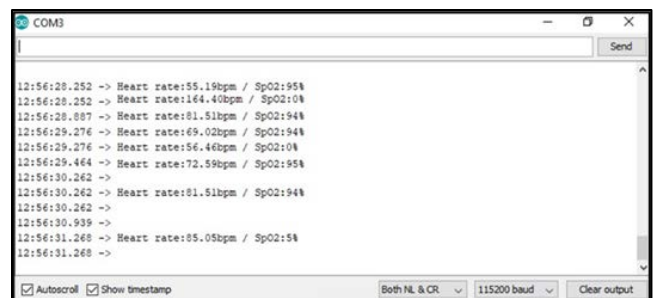

Fig. 8. Heartbeat Result Analysis.

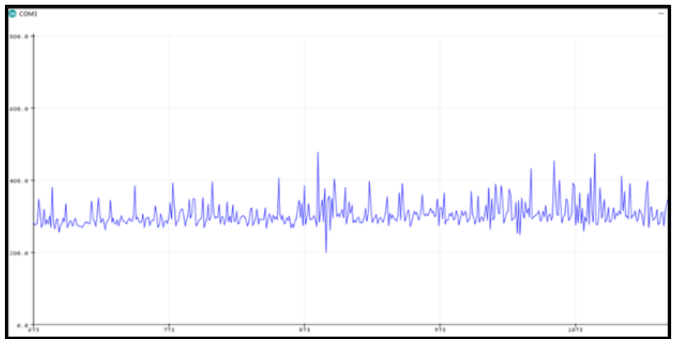

Fig. 9. Arduino com Port Result.

TABLE VIII. ECG_PQRST DATA FROM THE DEVICE

| Record No | Age | P | Q | R | S | T |
|---|---|---|---|---|---|---|
| 1 | 21 | 91.6 | 100 | 100 | 100 | 90 |
| 2 | 23 | 100 | 100 | 100 | 100 | 100 |
| 3 | 30 | 100 | 100 | 100 | 100 | 100 |
| 4 | 25 | 100 | 100 | 100 | 100 | 100 |
| 5 | 20 | 78.5 | 100 | 80 | 100 | 100 |
| 6 | 22 | 100 | 82.35 | 82.35 | 87.5 | 100 |
| 7 | 24 | 90 | 100 | 100 | 100 | 90 |
| 8 | 45 | 100 | 100 | 100 | 100 | 100 |
| 9 | 43 | 100 | 80 | 80 | 80 | 85 |
| 10 | 26 | 100 | 100 | 100 | 100 | 100 |
| 11 | 19 | 100 | 76.19 | 76.19 | 76.19 | 94.54 |
| 12 | 18 | 93.75 | 93.75 | 93.75 | 93.75 | 93.75 |
| 13 | 34 | 100 | 100 | 100 | 100 | 100 |
| 14 | 35 | 100 | 100 | 100 | 100 | 100 |
| 15 | 30 | 100 | 92.85 | 92.85 | 92.85 | 100 |
| 16 | 41 | 100 | 100 | 100 | 100 | 100 |
| 17 | 40 | 100 | 100 | 100 | 100 | 100 |
| 18 | 44 | 100 | 100 | 100 | 100 | 100 |
| 19 | 29 | 87.5 | 100 | 100 | 100 | 100 |
| 20 | 28 | 100 | 100 | 100 | 100 | 100 |

### H. ML (Machine Learning) Intercept and Coefficients

A statistical tool known as linear regression predicts the future value of heart condition measurement parameters (P, Q, R, S, T) by analyzing the past. The technique uses the least square method to draw a straight line that minimizes the difference between current values and resulting values. The best fit line associated with the n points (S1, T1), (S2, T2) ……., (Sn, Tn) has the form y = mx + b. Using the formula and we have got the intercept value which is 14.319164821638992.

From the analysis, the coefficient value for S is 0.962445, for T is -0.173491. For age, the value was 0.162937, where a positive sign implies that the response variable increases when the predictor variable rises. In contrast, a negative sign suggests that the response variable drops when the predictor variable increases.

### I. Prediction

The score of one variable is predicted based on the scores of a second variable. The criterion variable is the variable that forecasts and is denoted as Y. The variable on which the predictions are based is the predictor variable and abbreviated as X. The project used only one prediction variable: Y. The projections of Y are evaluated when it is plotted as a function of X from a straight line. This complete process of prediction is listed in the Table 9.

In Fig. 10 indicates a positive association between X and Y. If Y is forecasting based on X, the greater the value of X may provide an accurate prediction of Y. The regression line in the figure is made up of the expected score on Y for each possible value of X.

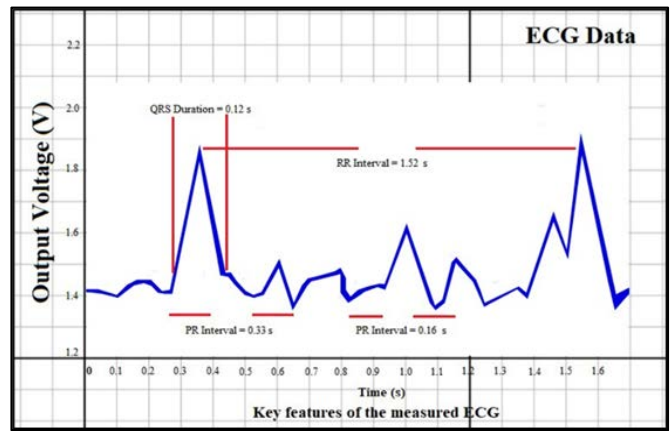

Fig. 10. The Key Feature of the Measured ECG.

TABLE IX. PREDICTION

| S. N | Actual | Predicted |
|---|---|---|
| 2 | 100 | 98.10267 |
| 5 | 82.35 | 84.76861 |
| 17 | 100 | 100.3838 |
| 19 | 100 | 97.7768 |
| 12 | 100 | 98.75442 |

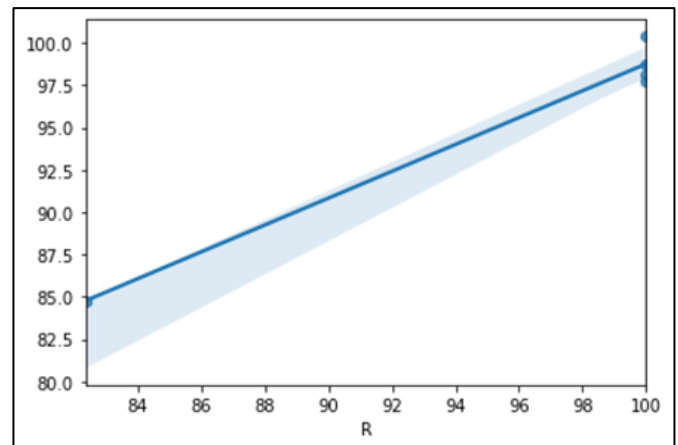

Fig. 11. Prediction Plot.

The error for a point is the difference between the actual value and the predicted value. From our analysis, the mean absolute error value is 1.633702533697678, and the mean squared error is 3.2181888486775514. Thus, the accuracy percentage is 93.5434261396096.

The bar diagram below (Fig. 12) shows the relationship between actual and predicted plots. For example, in Fig. 11, the blue indicates true, and the orange represents the expected value.

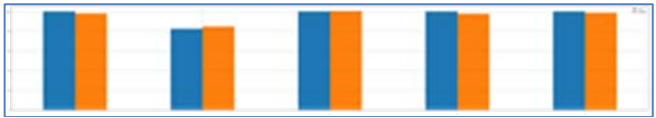

Fig. 12. Relationship between Actual and Predicted Data.

*J. Cost Analysis*

In developing countries, hospitals buy a lot of new medical equipment. Famous hospitals import medical equipment from other countries. As a result, hospital administrators pay a significant amount of money to bring the equipment to provide the best possible service to their patients.

The project described in this article is not costly. Every component is inexpensive and of good quality. As a result, it is affordable to everyone without causing financial hardship. The list of devices selected for ECG monitoring system purposes is listed in Table 10.

TABLE X. COST ANALYSIS

| Components | Unit Price |
|---|---|
| Arduino Mega 2560 | 800/- |
| Heartbeat/Pulse sensor (MAX30100) | 750/- |
| ECG sensor (AD8232) | 1250/- |
| Esp8266 A1 cloud Wi-Fi module | 180/- |
| CP2102 USB 2.0 to TTL UART Module | 180/- |
| Male-Female Jumper Wires | 100/- |
| 400 Tie Points Breadboard White | 100/- |
| Total | 3360/- |

## IV. CONCLUSION

We have created and executed an ECG monitoring system that is entirely based on current IoT technologies. The IoT-based ECG monitoring system is constructed based on the proposed design. IoT-based healthcare platform links with smart sensors affixed to the human body for health monitoring. We talked about IoT-based patient monitoring systems in this article. Smartphones or gadgets use intelligent technologies, and we have discussed the advantages, disadvantages, and opportunities. Continuous remote monitoring is required for observing the medical patient. Our research work provides the ability to monitor patients via web app services and mobile massage services continuously. This research also contrasted the early medical system to modern health monitoring. The work will bring change in medical science and be a blessing for rural areas. The research work has proved its benefits already. We are planning for the further development of the project by promising that one day every people of our country will get immediate medical treatment with the help of our project.